\documentclass[twocolumn]{aastex631}
\usepackage{apjfonts, natbib}
\usepackage{mathtools}

\newcommand{\lum}{erg\,s\ensuremath{^{-1}}}

\newcommand{\nii}{N\,{\footnotesize II}}

\newcommand{\oiii}{[O\,{\footnotesize III}]}

\newcommand{\sersic}{S\'{e}rsic}

\begin{document}

\title{Embers of Active Galactic Nuclei:  Tidal Disruption Events and Quasiperiodic Eruptions}

\author[0000-0002-7152-3621]{Ning Jiang}
\email{jnac@ustc.edu.cn}
\affiliation{CAS Key Laboratory for Research in Galaxies and Cosmology, Department of Astronomy, University of Science and Technology of China, Hefei 230026, People’s Republic of China}
\affiliation{School of Astronomy and Space Sciences, University of Science and Technology of China, Hefei 230026, People’s Republic of China}

\author[0000-0001-9608-009X]{Zhen Pan}
\email{zhpan@sjtu.edu.cn}
\affiliation{Tsung-Dao Lee Institute, Shanghai Jiao-Tong University, Shanghai, 520 Shengrong Road, 201210, People’s Republic of China}
\affiliation{School of Physics \& Astronomy, Shanghai Jiao-Tong University, Shanghai, 800 Dongchuan Road, 200240, People’s Republic of China}
\affiliation{Tsung-Dao Lee Institute, School of Physics and Astronomy, Shanghai Jiao Tong University,
Key Laboratory for Particle Physics, Astrophysics and Cosmology (Ministry of Education),
Shanghai Key Laboratory for Particle Physics and Cosmology,
800 Dongchuan Road, Shanghai, 200240, P.R. China}

\begin{abstract}
Recent observations have confirmed the direct association between tidal disruption events (TDEs) and quasiperiodic eruptions (QPEs). In addition, TDE hosts and QPE hosts are statistically found to be similar in their  morphological properties and in the strong overrepresentation of poststarburst galaxies. Particularly, both of them show an intriguing preference for extending emission line regions (EELRs), which are indicative of recently faded active galactic nuclei (AGNs). 
This further suggests that QPEs might be produced following TDEs involving supermassive black holes at a particular stage, when the AGN activity has recently ceased.
Moreover, in the framework of “QPEs=extreme mass ratio inspiral (EMRI) + accretion disk" model, a large fraction of QPE EMRIs are inferred to be quasi-circular from the QPE timing, indicating that they are wet EMRIs that were formed in the AGN disk during a previous AGN phase. Based on these facts, we propose a unified scenario that connects these three phenomena: AGN activities boost both the TDE rate and the formation rate of low-eccentricity EMRIs,
consequently TDEs are preferentially found in recently faded AGNs instead of in ongoing AGNs due to selection effects, 
and QPEs are also  preferentially found in recently faded AGNs where TDEs frequently feed a misaligned accretion disk to the EMRI.
\end{abstract}



\section{Introduction} \label{sec:intro}

In the past decade, advancements in survey technologies across various wavelengths have resulted in a fast increasing detection rate of rare transient events associated with supermassive black holes (SMBHs) located in the centers of galaxies. They include stellar tidal disruption events (TDEs, see \citealt{Gezari2021} as a review), changing-look active galactic nuclei (AGNs, \citealt{LaMassa2015}) and X-ray quasiperiodic eruptions (QPEs, \citealt{Miniutti2019GSN069}). Among them the physical origin of QPEs is particularly puzzling due to their fascinating observational properties and rarity, 
with about 10 QPE sources having been reported so far \citep{Sun2013,Miniutti2019GSN069,Giustini2020RXJ1301,Arcodia2021,Arcodia2022,Arcodia2024,Chakraborty2021,Evans2023,Guolo2024,Nicholl2024Natur}. A significant progress very recently is the direct detection of QPEs following a confirmed optical TDE AT2019qiz \citep{Nicholl2024Natur}, indicating that at least some (if not all) QPEs are firmly connected with TDEs. Such a connection was previously tentatively observed in the first QPE source GSN~069~\citep{Miniutti2019GSN069} although it is not a standard TDE with optical spectroscopic identification yet with evidence in its long-term X-ray light curve~\citep{Shu2018} and abnormal nitrogen-enriched gas implied by its UV and X-ray spectra~\citep{Sheng2021,Kosec2025}. In addition, the QPE-TDE association also appears in some other sources prior to AT~2019qiz, although it is not as conclusive due to a lack of clear confirmation of QPE, in the case of AT~2019vcb~\citep{Quintin2023}, or TDE, in the case of XMMSL1~J024916.6-041244~\citep{Chakraborty2021}. Additional eruptions of AT~2019vcb reported by \cite{Bykov2024} further increase its credibility as a QPE source.

Further insights can be gained from the resemblance in host galaxy characteristics between TDEs and QPEs. TDEs are found to be strongly overrepresented in low-mass poststarburst (PSB) or E+A galaxies~\citep{Arcavi2014,French2016,Law-Smith2017} which show a highly concentrated bulge with high \sersic\ indices, bulge-to-total light (B/T) ratios, and stellar surface mass densities compared to the overall galaxy population~\citep{French2020}. Interestingly, a similar preference of host galaxies has been also observed in QPEs recently~\citep{Wevers2022,Gilbert2024}.
In addition, \citet{Wevers2022} performed a systematic study of the host galaxies of five known QPE sources using new and archived medium-resolution optical spectra and claimed that all QPE host galaxies have emission lines indicating the presence of an AGN. Further integral field spectroscopic observations suggest a high incidence (3/5) of extended emission line region (EELR) detection in the hosts of QPEs~\citep{Wevers2024-QPE}. These EELRs extend up to 10~kpc in size, requiring photoionization by a nonstellar continuum, but far beyond the current AGN luminosity. 
In a companion study, \citet{Wevers2024-TDE} estimated the incidence of EELRs in PSB-TDE hosts and found it is a factor of $\sim10\times$ higher than in other PSB galaxies, suggesting that a gas-rich postmerger environment is a key ingredient in driving elevated TDE rates. Therefore, both TDEs and QPEs actually show a preference for a specific evolutionary stage of PSB galaxies, namely the phase of recently faded AGNs~\citep{French2023}.

The growing evidence of TDE-QPE association promoted a ``TDE+EMRI=QPEs" model \citep{Linial2023},
where the QPEs are the result of quasiperiodic collisions between an extreme mass ratio inspiral (EMRI) and an accretion disk formed from a previous TDE.
This model or the general EMRI+accretion disk model \citep{Franchini2023,Tagawa:2023fpb} naturally explains the intriguing features in the QPE timing, including  the alternating long-short $T_{\rm long} -  T_{\rm short}$ pattern  found in recurrence times of several QPE sources \citep{Miniutti2019GSN069,Giustini2020RXJ1301,Arcodia2021,Arcodia2022,Miniutti2025}, 
and an approximately constant sum of two consecutive recurrence times $T_{\rm long}+ T_{\rm short}$ with both $T_{\rm long}$ and $T_{\rm short}$ showing large variations  with time \citep{Zhou2024a}. 
In the framework of EMRI+accretion disk model, previous studies have demonstrated the feasibility of measuring  
the SMBH mass and the EMRI orbital parameters from the QPE timing \citep{Xian2021,Franchini2023,Zhou2024a,Zhou2024b,Zhou2024c}.
Systematic analyses show that EMRIs that are sourcing QPEs are with orbital radii of $\mathcal{O}(10^2) M_\bullet$,
and a large fraction of them are of  low orbital eccentricity with $e=\mathcal{O}(10^{-2})$ \citep{Zhou2024a, Zhou2024b,Zhou2024c},
where $M_\bullet$ is the gravitational radius of the SMBH.
These low-eccentricity QPE EMRIs can be naturally interpreted as wet EMRIs formed in accretion disks of previous AGNs \citep{Sigl2007,Levin2007,Pan2021prd,Pan2021b,Pan2021,Pan2022,Derdzinski2023}, 
but are too circular to be compatible with predictions of  loss-cone channel \citep{Hopman2005,Preto2010,Bar-Or2016,Babak2017,Amaro2018,Broggi2022} or Hills channel \citep{Miller2005,Raveh2021}. Therefore, past AGN activities are inevitable for providing the necessary condition (a low-eccentricity EMRI) for a large fraction of QPEs.

Based on these observations and theoretical analyses,  we propose a unified scenario connecting the three phenomena: 
AGN activities boost both the TDE rate and the EMRI formation rate, and consequently the QPE rate.  
In Section~\ref{sec:data}, we review the QPE observations and emphasize the evidence for their associations with TDEs and recently faded AGNs.
In Section~\ref{sec:EMRIs}, we summarize the population properties of QPE sources and their connections to past AGN activities. 
In Section~\ref{sec:conclusion}, we elaborate the unified scenario that explains the associations between AGNs, TDEs and QPEs. 
In this Letter, we will use geometric units with $G=c=1$ if not specified otherwise.



\section{QPE Sample}
\label{sec:data}

We briefly review each of the QPE sources reported in the literature, with a particular focus on their evidence for associations with TDEs and recently faded AGNs, if such associations exist. A total of 9 QPEs or high-confidence QPE candidates have been considered in this work. The repeating soft X-ray outbursts in Swift~J023017.0+283603 has been also proposed as possible QPE by some works~\citep{Evans2023,Guolo2024}, but its much longer recurrent timescale ($\sim21$ days) and its opposite behaviours in the shape and spectral evolution during the flares do not allow a confident association with QPEs for the time being. We have thus not included it in our sample.

The TDE evidence can be either supported by a standard optical identification, i.e. the selection criteria adopted by the Zwicky Transient Facility~\citep{vV2021,Hammerstein2023}, or by long-term X-ray light curves characterized by an soft X-ray outburst~\citep{Saxton2020}. Here, the recently faded AGN means that the gas accretion of the SMBH has just turned off, leading to the disappearance of AGN continuum and broad-line emissions.  However, it still leaves some imprints of the historic activity, i.e. a narrow-line region (or EELR) photoionized by nonstellar (AGN) continuum or a bright IR echoes indicating residual torus. In the context of changing-look AGNs, a type 2 AGN can be actually a recent turn-off AGN rather than being viewed from edged-on~\citep{Sheng2017,Hutsemekers2017}. This is especially true if a TDE or a QPE has been observed in a type 2 AGN, which is otherwise blocked by the torus and thus unseen in the canonical picture of AGN unification~\citep{Antonucci1993}.

\begin{deluxetable*}{llcccc}
\tablenum{1}
\tablecaption{The basic informations of QPEs and their associations with TDE and recently faded AGN  \label{tab:QPEsample}}
\tablewidth{0pt}
\tablehead{
\colhead{Source}  &  \colhead{Redshift} & \colhead{Black hole mass $\log_{10}(M_\bullet/M_\odot)$ } & \colhead{TDE} & \colhead{Recently faded AGN } & \colhead{References} }
\decimalcolnumbers
\startdata
RXJ1301  & 0.0237 & $6.65\pm 0.42$ & No & Yes & 1,4  \\
GSN~069  & 0.0182 & $5.99\pm0.5$ & Yes & Yes & 1,4,5,6 \\
AT~2019qiz & 0.0151 & $6.48\pm0.33$  & Yes  & Yes & 2,7,8 \\
eRO-QPE1 & 0.0505 & $5.78\pm0.55$ & ?  & No? & 1,4,9 \\
eRO-QPE2 & 0.0175 & $4.96\pm0.54$ & ? &  Yes & 1,4,9 \\
eRO-QPE3 & 0.024 & $5.1\pm0.55$ &  ? & No? & 3,4 \\
eRO-QPE4 & 0.0437 & $7.24^{+0.32}_{-0.12}$ & ? & ? & 3 \\
J0249	& 0.0186 &   $5.29\pm0.55$ & Yes  & ? & 1,10 \\
AT~2019vcb & 0.089 & $6.03\pm0.39$ & Yes & ? & 2,11,12 \\
\enddata
\tablecomments{
Columns (1)-(3): the name, redshift and black hole mass in units of $\log_{10}(M_\bullet/M_\odot)$ of the QPE sources. Columns (4)-(5): Whether or not the QPE is associated with a TDE or a recently faded AGN. Here "?" means that there is no clear evidence yet, but this may be due to a lack of sufficient observations.  Column (6): Related references: 1~\citep{Wevers2022}; 2~\citep{Yao2023}; 3~\citep{Arcodia2024}; 4~\citep{Wevers2024-QPE}; 5~\citep{Shu2018}; 6~\citep{Sheng2017}; 7~\citep{Nicholl2024Natur}; 8~\citep{Xiong2025}; 9~\citep{Arcodia2021}; 10~\citep{Chakraborty2021}; 11~\citep{Quintin2023}; 12~\citep{Bykov2024}.
Note the black hole mass measurements in column (3) depend on the scaling relation used, e.g., a different $M_\bullet-\sigma_\star$ relation used in \cite{Wevers2019} prefers a lower value of J0249 black hole mass $\log_{10}(M_\bullet/M_\odot)=4.93^{+0.55}_{-0.53}$. Also we noticed that the mass of eRO-QPE4 might be overestimated systematically.}
\end{deluxetable*}

\begin{itemize}

  \item {\bf RXJ1301}: This is the first proposed QPE candidate which displays X-ray flares in both archival XMM-Newton and Chandra observations while being discovered as a supersoft AGN~\citep{Sun2013}. Its QPE nature was then confirmed by \citet{Giustini2020RXJ1301}. Its optical spectrum taken in 2007 by the Sloan Digital Sky Survey (SDSS) was dominated by narrow lines after continuum subtraction, the line ratios of which place it in the Seyfert regime on the Baldwin-Phillips-Terlevich (BPT) diagram of~\citet{Kewley2006}.
  An EELR with the largest (projected) radius ($R_{\rm max}$) to the galactic nucleus  $\sim 3$ kpc has been discovered~\citep{Wevers2024-QPE} in the observations of Multi Unit Spectroscopic Explorer (MUSE) integral field spectroscopy mounted on very large telescope (VLT). The EELRs are photoionized by a nonstellar continuum, but the current nuclear luminosity is insufficient to power the observed emission lines, which requires a minimum nuclear ionizing luminosity ($L_{\rm ion,min}$) of $2.5\times10^{42}$~\lum\ to power the EELR~\citep{Wevers2024-QPE}.

  \item {\bf GSN~069}: This is the first confirmed QPE source~\citep{Miniutti2019GSN069}. It was initially reported as a true Seyfert 2 galaxy candidate since a bright soft X-ray source  was detected  at a position consistent with it in 2010, but its optical spectra are always dominated by unresolved emission lines with no broad components~\citep{Miniutti2013}. Its nondetection by ROSAT implies it has undergone X-ray brightening by a factor of $\gtrsim 240 $, suggesting a likely TDE, which has been supported by its subsequent long-term slow decay~\citep{Shu2018}. The VLT/MUSE observation reveals a EELR with $R_{\rm max}\sim9$ kpc, which calls for a 
  $L_{\rm ion,min}$ of $6\times10^{42}$~\lum~\citep{Wevers2024-QPE}. It is also notable that \citet{Patra2024} found evidence of a compact nuclear \oiii\ region making use of archival HST data, which indicates that the accretion activity is only $\sim10-100$ yr old.
  
  \item {\bf AT2019qiz}: This is the first optical TDE with unambiguous QPE detection~\citep{Nicholl2024Natur}. The ratios of narrow lines in the BPT diagram lie intermediate between the region of star-forming galaxies and AGNs, indicative of a possible weak AGN prior to the TDE~\citep{Nicholl2020}. After a careful study of the VLT/MUSE data of its host, we found an EELR, which is most prominent in the \nii\ map with scale of $\sim3.7$ kpc. It provides further evidence of AGN ionizing in the near past~\citep{Xiong2025}. The detection of a prominent IR echo~\citep{Short2023} also suggests a likely residual dusty torus of a recently faded AGN. In contrast, the echoes of normal TDEs are rather weak due to a much lower dust covering factor~\citep{Jiang2021}.

   \item {\bf eRO-QPE1}: This is one of the first two, together with eRO-QPE2, QPE sources discovered in galaxies without obvious signatures of AGNs~\citep{Arcodia2021}. However, the high-quality Magellan/MagE spectroscopy suggests clear AGN activity based on the BPT diagnosis~\citep{Wevers2022}. It has also been observed by VLT/MUSE, but no EELR has been detected, possibly due to its higher redshift and/or a fainter and more compact host galaxy~\citep{Wevers2024-QPE}.

   \item {\bf eRO-QPE2}: The discovery paper claims that its optical spectra shows intense narrow emission lines mainly contributed by strong star-forming activity~\citep{Arcodia2021} while the high-quality Magellan spectroscopy places it in the Seyfert AGN region~\citet{Wevers2022} based on the selection criteria presented in~\citet{Cid2011}. The VLT/MUSE observation reveals an EELR with $R_{\rm max}\sim4.6$~kpc and a $L_{\rm ion,min}$ of $6\times10^{42}$~\lum~\citep{Wevers2024-QPE}.

   \item {\bf eRO-QPE3}: This is the third QPE discovered by eROSITA, there are no strong canonical signatures of AGN from current optical, UV, infrared and radio data~\citep{Arcodia2024}. The SALT spectrum shows no apparent emission lines while weak narrow lines are detectable on the continuum-subtracted spectrum. Although a precise diagnosis is difficult, a strong preexisting AGN is disfavored. 

   \item {\bf eRO-QPE4}: Its quiescent X-ray emission was a few times fainter, or absent before the QPE discovery, supporting a transient nature for the quiescent accretion disk and disfavoring a long-lived radiatively efficient accretion flow~\citep{Arcodia2024}. Similar to eRO-QPE3, only very faint narrow lines are visible in the SALT spectrum even after continuum subtraction.

   \item {\bf XMMSL1~J024916.6-041244 (hereafter J0249)}: This is the fifth QPE candidate with 1.5 QPE-like soft X-ray flares discovered by a blind search of the XMM-Newton Source Catalog~\citep{Chakraborty2021}.
   It was first identified as a TDE candidate based on a flux increase by a factor of 88 compared to a ROSAT upper limit, and a very soft X-ray spectrum that can be described by a blackbody~\citep{Strotjohann2016}. Its host galaxy was initially classified as a Seyfert 1.9~\citep{Esquej2007}, which has recently been reclassified as a star-forming galaxy potentially hosting an AGN by higher resolution spectra~\citep{Wevers2019,Wevers2022}.

   \item {\bf AT 2019vcb (aka Tormund)}: This is an optical TDE discovered by the ZTF~\citep{Hammerstein2023}. Its hint of QPE was first found during an archival search for QPEs in the XMM-Newton archive~\citep{Quintin2023}. The eROSITA happened to observe the source 13 days after the XMM-Newton observation, that is seven months after the onset of the optical TDE, and detected the other flare~\citep{Bykov2024}. Both flares bear similar characteristics in terms of timing and spectral properties, providing additional evidence of QPE. However, its QPE nature is not as conclusive as other sources, as only two isolated flares have been detected, thus we treat it as a QPE candidate for safety.
   Its host galaxy does not show any signature of AGN~\citep{Quintin2023}.
\end{itemize}

In summary, among the 9 known QPEs or high confidence QPE candidates, two (AT2019qiz, AT2019vcb) were detected after standard optical TDEs and the other two (GSN~069, J0249) were detected during the declining phase of likely X-ray TDE candidates. Additionally, four (RXJ1301, GSN069, AT2019qiz, eRO-QPE2) exhibit clear EELRs in the  integral field unit (IFU) observations, indicating the presence of recently faded AGNs. Quantitatively, the typical kpc physical scale indicates that the SMBHs were actively accreting, formally in the AGN phase, at least tens of thousands of years ago. It is worth noting that the four showing EELRs happen to be among the five lowest redshifts ($z<0.024$, see Table~\ref{tab:QPEsample}) and the only exception is J0249 at $z=0.0186$ which has no IFU observation to date. This strongly suggests that the nondetection of other sources (e.g., eRO-QPE1 and eRO-QPE3) could be a selection effect as the detection of EELRs in more distant targets is obviously more challenging and requires a higher sensitivity and spatial resolution of the observations. 
Notably, AT2019qiz, the first unambiguous QPE in a spectroscopically confirmed TDE, and GSN 069, the first reported QPE in a likely X-ray TDE candidate, 
both show detections of EELRs, further supporting that QPEs, TDEs and recently faded AGNs are all closely related.

\begin{deluxetable*}{llccc}
\tablenum{2}
\tablecaption{EMRI orbital properties of QPE sources  \label{tab:QPE-EMRIs}}
\tablewidth{0pt}
\tablehead{
\colhead{Source}  &  \colhead{Orbital period $T_{\rm obt}\left(\approx 2\langle{T}_{\rm rec}\rangle\right) $ [ks]}  & \colhead{${T}_{\rm obt} \propto M_\bullet^{0.8}$}
& \colhead{Orbital eccentricity $e$} & \colhead{EMRI formation (compatible w. dry/Hills channels)}}
\decimalcolnumbers
\startdata
RXJ1301  &  $32$~\citep{Giustini2020RXJ1301} & No &  $0.25^{+0.18}_{-0.20}$ & Yes \\
GSN~069  &  $65$~\citep{Miniutti2019GSN069} & Yes & $0.04^{+0.02}_{-0.03}$ &  No\\
AT~2019qiz & $340$~\citep{Nicholl2024Natur} & Yes & $\sim 0.13$ & No?\\
eRO-QPE1 &  $140$~\citep{Arcodia2021} &  Yes & $0.10^{+0.47}_{-0.10}$& No\\
eRO-QPE2 &  $17$~\citep{Arcodia2021} &  Yes & $0.01^{+0.02}_{-0.01}$ & No \\
eRO-QPE3 &  $150$~\citep{Arcodia2024} &  Yes & $\sim 0$ & No\\
eRO-QPE4 &  $90$~\citep{Arcodia2024} & No & $\sim 0.18$ & Yes\\
J0249	&  $19$~\citep{Chakraborty2021}& Yes & ? & ?\\
AT~2019vcb & $\sim 110 - 500$~\citep{Bykov2024}  & Yes  & ? & ?\\
\enddata
\tablecomments{In column (2), the orbital period $T_{\rm obt}$ is identified as twice the average recurrence time $\langle{T}_{\rm rec}\rangle$ of QPEs.
The QPE recurrence time $\langle{T}_{\rm rec}\rangle$ of AT~2019vcb is most uncertain since no consecutive flares have been detected. Assuming the flare duration of $\sim 10 - 20$ hours \citep{Bykov2024}, and the tentative correlation between recurrence time and flare duration shown in Fig.~3 of \cite{Nicholl2024Natur}, we find $\langle{T}_{\rm rec}\rangle\sim 15 - 70$ hours.  
In column (3), the information of $T_{\rm obt}-M_\bullet$ correlation is read from  Fig.~3 of \cite{Zhou2024c}. In  column (4), orbital eccentricities (95\% confidence level)  of  RXJ1301, GSN 069, eRO-QPE1 and eRO-QPE2 are obtained by full orbital analyses of the QPE timing data \citep{Zhou2024b,Zhou2024c},
while orbital eccentricities of AT~2019qiz, eRO-QPE3 and 4 are roughly estimated using Eq.~(\ref{eq:e_est}), since the numbers of QPEs available are too low for reasonable orbital analyses. Only a few flares in 
J0249 and  AT~2019vcb have been detected, from which no orbital eccentricity information can be extracted.
In column (5), the EMRI formation channel information is based on analyses by \cite{Zhou2024b}: the two most eccentric ones are
compatible with dry/Hills channels, 
four QPE EMRIs are incompatible with dry/Hills channels but are consistent with the wet channel prediction, and more observations are needed for analyzing the remaining three.
}
\end{deluxetable*}

\section{Orbital properties and formation processes of QPE EMRIs}
\label{sec:EMRIs}

Though a number of models have been proposed for explaining the physical origin of QPEs in the past years,
more and more theoretical analyses and observations favor the EMRI+accretion disk model as the physical origin of QPEs \citep{Linial2023,Franchini2023,Tagawa:2023fpb}.
In this framework, 
the central SMBH mass and the EMRI orbital parameters can be extracted from the QPE timing data as shown in previous 
studies \citep{Xian2021,Franchini2023,Zhou2024a,Zhou2024b,Zhou2024c}, based on which one can further infer the EMRI formation channels \citep{Zhou2024a,Zhou2024b},
including (wet) AGN disk channel,  (dry) loss-cone channel and Hills channel.

In the AGN disk channel \citep{Sigl2007,Levin2007,Pan2021prd,Pan2021b,Pan2021,Pan2022,Derdzinski2023}, a stellar mass object (SMO) which was either born in the disk or captured into the disk, is expected to migrate inward driven by density waves excited in the disk until $r=\mathcal{O}(10^2) M_\bullet$, where gravitational wave (GW) emissions 
become the dominant driving force. EMRIs formed in AGN disks are expected to be of low eccentricity due to  efficient eccentricity damping by the density waves.
In the loss-cone channel \citep{Hopman2005,Preto2010,Bar-Or2016,Babak2017,Amaro2018,Broggi2022},  a SMO in the nuclear stellar cluster 
is randomly scattered into a low angular momentum orbit where GW emissions dominate over two-body scatterings from other SMOs in the nuclear cluster,  
then the orbit shrinks driven by GW emissions.
EMRIs formed in the loss-cone channel are expected to keep a large residual orbital eccentricity.
In the Hills channel \citep{Miller2005}, a binary of SMOs in the nuclear cluster is disrupted by the SMBH, with one SMO ejected and the other captured by the SMBH on an eccentric 
orbit. The ensuing orbital evolution depends on whether the captured SMO is in the GW emission dominated regime or in the two-body scattering dominated regime. 
Numerical simulations \citep{Raveh2021} show that Hills EMRIs are similar to loss-cone EMRIs in the residual orbital eccentricity. 

Detailed orbital analyses and EMRI formation inference have been conducted by \cite{Zhou2024a,Zhou2024b,Zhou2024c} 
for 4 QPE sources (RXJ1301, GSN 069, eRO-QPE 1 and 2)  where a reasonable number of eruptions are available (see  Table~\ref{tab:QPE-EMRIs}).
\cite{Zhou2024b} found that GSN 069, eRO-QPE 1 and 2 are beyond the reach of dry/Hills channels in the $(A, r_{\rm p})$ phase space, even taking the uncertainties in the semi-major axis $A$
and the eccentricity $e$ into account, where  $r_{\rm p}= A(1-e)$ is the pericenter distance.
It is interesting to find that their orbital parameters are consistent with the wet channel prediction.  
\cite{Zhou2024b} also found RXJ1301 is compatible with dry/Hills channels.

No orbital analysis has been done for AT 2019 qiz or eRO-QPE 3, 4 due to a limited number of eruptions available. We therefore roughly estimate their EMRI orbital eccentricities
by \citep{Pasham:2024sox}
\begin{equation}\label{eq:e_est}
    e\approx \frac{\pi}{4} \frac{T_{\rm long}-T_{\rm short}}{T_{\rm long}+T_{\rm short}}\ ,
\end{equation}
and orbital periods by twice the average recurrence times, $T_{\rm obt}\approx 2\langle{T}_{\rm rec}\rangle$, where $T_{\rm long}, T_{\rm short}$ are the consecutive long and short recurrence times. 
We then examine whether the QPE EMRIs are compatible with dry/Hills channel in $(A, r_{\rm p})$ phase space as done by \cite{Zhou2024b}.
As a result, we find the eRO-QPE4 EMRI with $\log_{10}(M_\bullet/M_\odot) = 7.24^{+0.32}_{-0.12}, T_{\rm obt}= 90$ ks, $e\sim 0.18$  is compatible with dry/Hills channels. In the similar way, we find eRO-QPE3 EMRI is incompatible with dry/Hills channels  
for any reasonable eccentricity, say $e < 0.3$, that is possibly consistent with the $T_{\rm long}\approx T_{\rm short}$ pattern in the recurrence times. The AT 2019 qiz EMRI is found to be incompatible with dry/Hills channels even assuming a large orbital eccentricity $e=0.26$ that is twice the rough estimation $e\sim 0.13$ (we put a question mark in its classification in Table~\ref{tab:QPE-EMRIs} for safety). 

In short, 4 QPE EMRIs (GSN 069 and eRO-QPE1, 2, 3) are  incompatible with dry/Hills channels,  but are consistent with the wet EMRI prediction, therefore are likely wet EMRIs that were formed in accretion disks of previous AGNs;
2 most eccentric QPE EMRIs (RXJ1301 and eRO-QPE4) are compatible with dry/Hills channels, 
which were possibly formed either in a quiescent phase of the SMBH or as misaligned EMRIs in an AGN phase.
For classifying the remaining three (AT~2019qiz, J0249, AT~2019vcb) EMRIs, more follow-up observations are needed for pinning down their orbital parameters and the SMBH masses.

As a piece of indirect evidence for multiple EMRI formation channels contributing to QPE EMRIs, \cite{Zhou2024c} found a likely 
correlation between the EMRI orbital period and the SMBH mass 
$T_{\rm obt}\propto M_\bullet^{0.8}$ for  the majority of  QPE EMRIs, except  the two most eccentric ones (RXJ1301 and eRO-QPE4). 
This dichotomy indicates that a large fraction of QPE EMRIs were formed in the same  channel  which enables the
$T_{\rm obt}- M_\bullet$ correlation, and the remaining EMRIs were formed in different channels,
which generate either a different $T_{\rm obt}- M_\bullet$ correlation or no $T_{\rm obt}- M_\bullet$ correlation at all,
depending on the details of  physical processes in these channels.

In summary, among the 9 known QPE EMRIs, four of them (GSN 069 and eRO-QPE1,2,3) are incompatible with dry/Hills channels
and are likely wet EMRIs, while the two most eccentric ones (RXJ1301 and eRO-QPE4)
  are compatible with dry/Hills channels. Moreover, the majority of the  QPE EMRIs are found to follow a  likely $T_{\rm obt}-M_\bullet$ correlation, except the two most eccentric ones.
 It is interesting to note that all the four likely wet EMRIs happen to follow the likely $T_{\rm obt}-M_\bullet$ correlation, 
suggesting a wet origin of this correlation. 
 These observations and EMRI formation modeling show that a previous AGN phase is inevitable for providing the necessary condition (a low-eccentricity EMRI) for a large fraction ($\gtrsim 4/9$) of QPE sources.
As shown in Section~\ref{sec:data}, GSN 069 and eRO-QPE2 are indeed found exhibiting clear EELRs, indicating the presence of recently faded AGNs.
The strong preference of detecting EELRs in low-redshift sources indicates that 
the nondetection in  higher-redshift QPE (e.g., eRO-QPE1 and eRO-QPE3) hosts  is possibly due to a selection effect.
 
\begin{figure*}
\centering
\begin{minipage}{1.0\textwidth}
\includegraphics[width=0.5\textwidth]{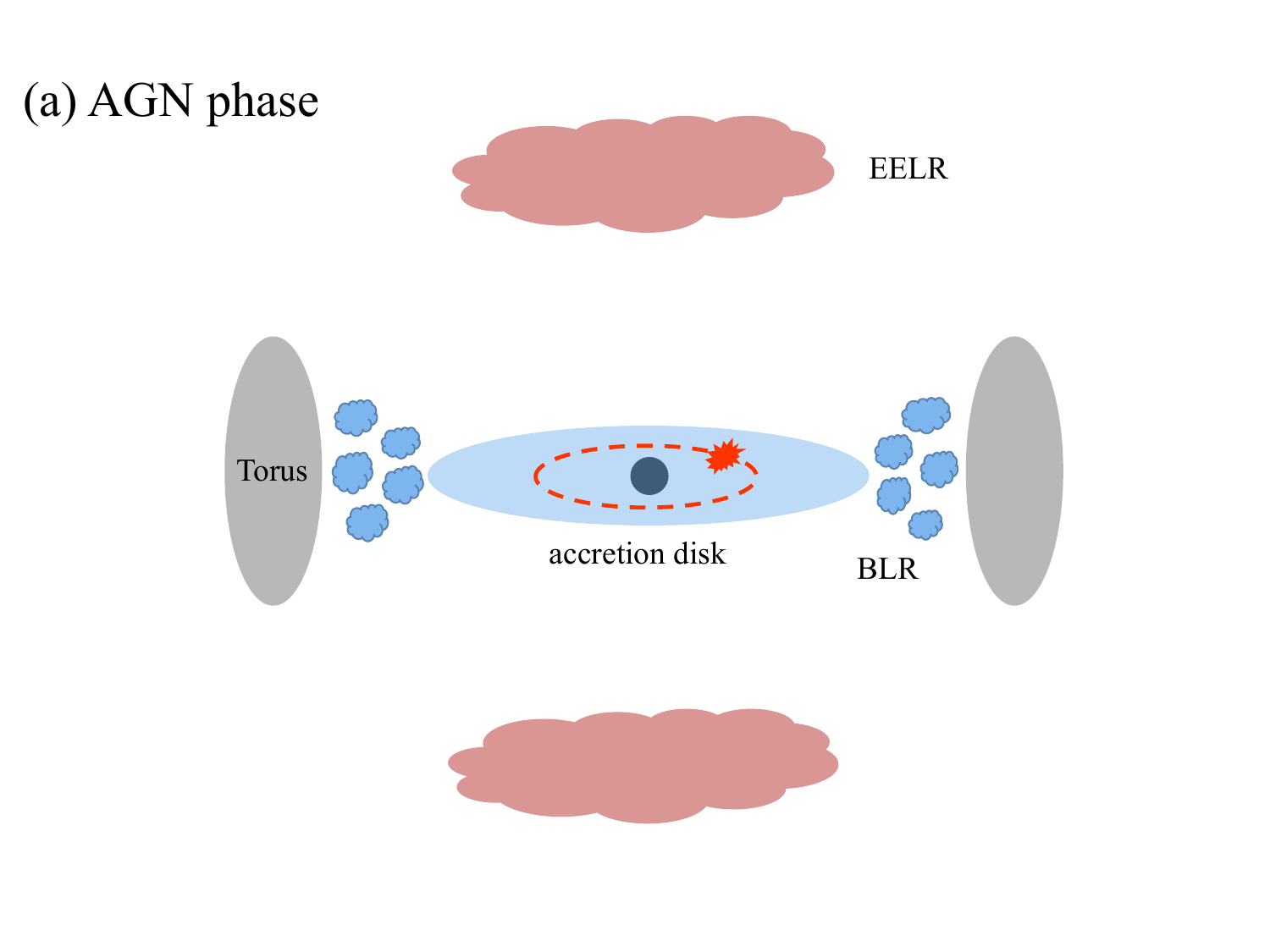}
\includegraphics[width=0.5\textwidth]{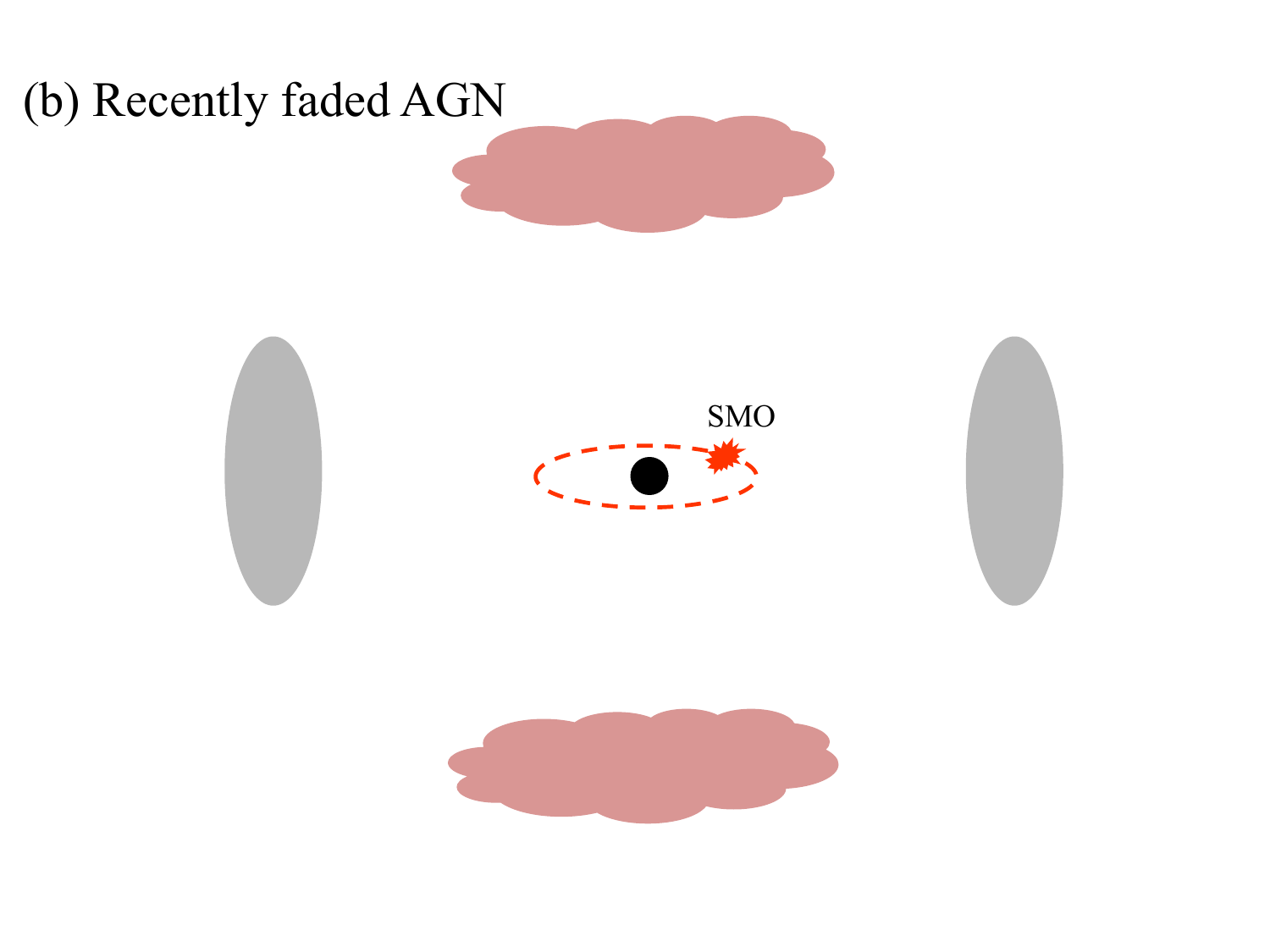}
\includegraphics[width=0.5\textwidth]{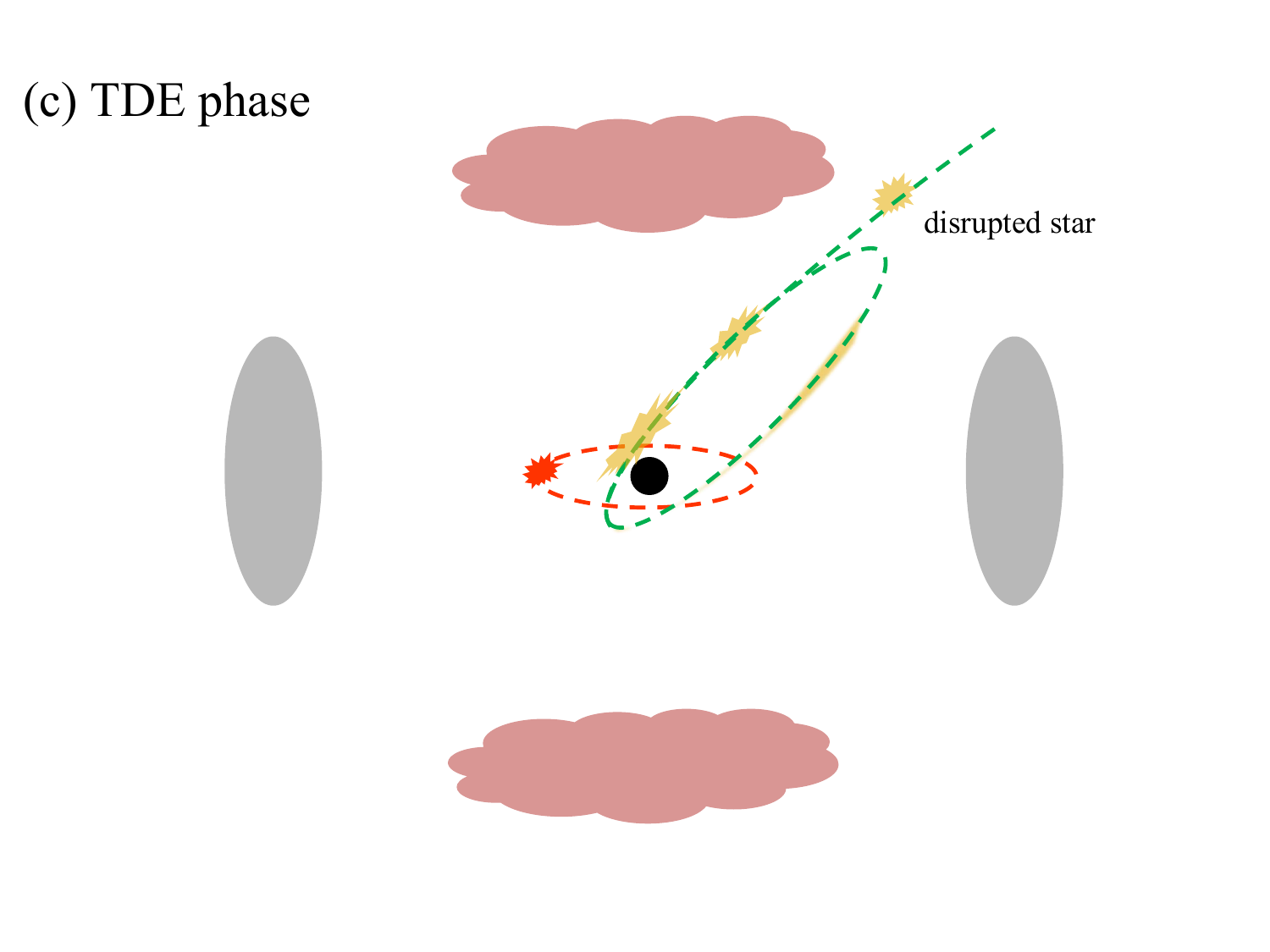}
\includegraphics[width=0.5\textwidth]{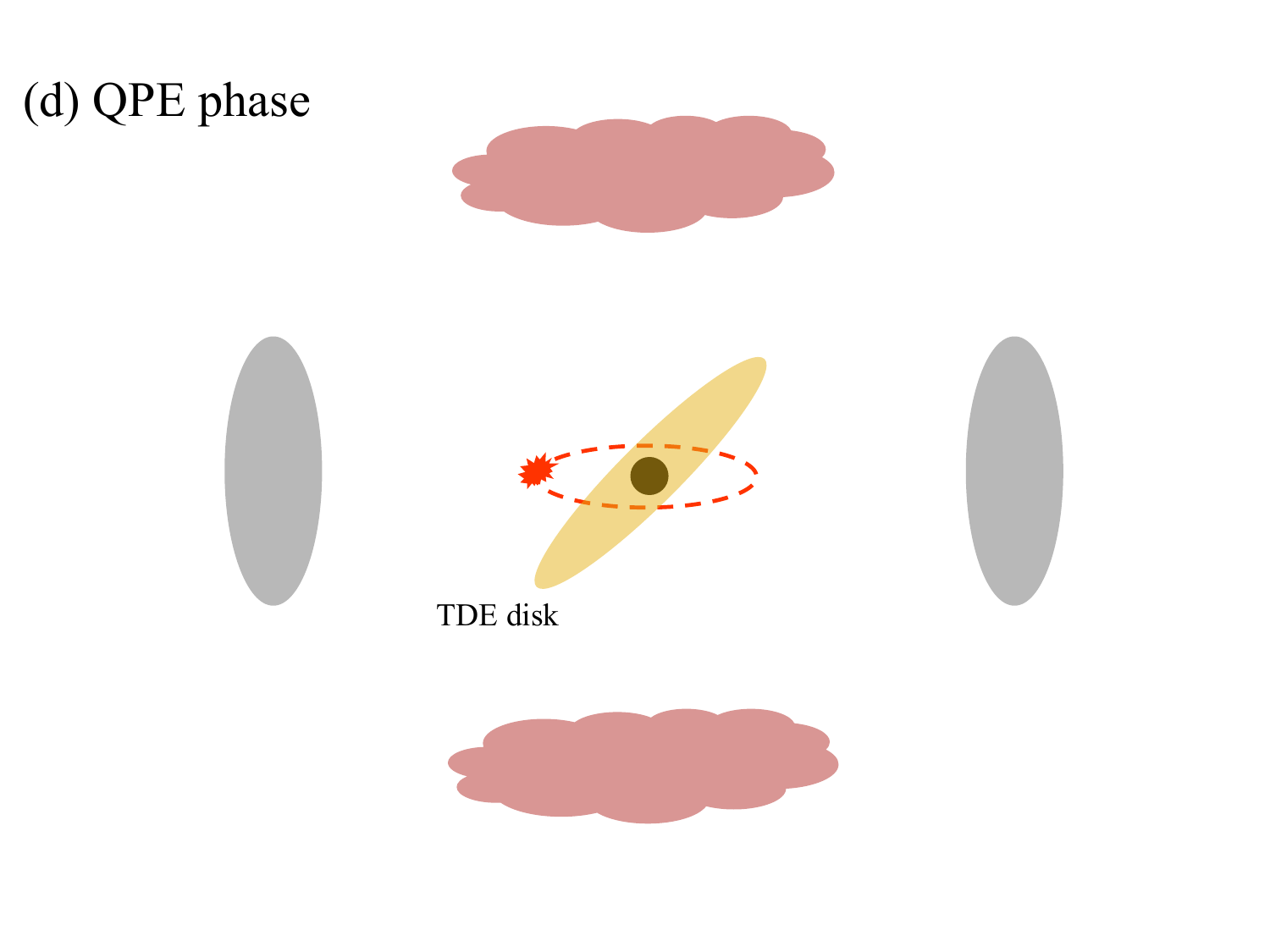}
\end{minipage}
\caption{Schematic picture of TDEs and QPEs as embers of AGNs. Note that the scale does not reflect the actual sizes, which are otherwise too small to visualize for the central part.
Panel (a): The AGN phase, in which we have shown the standard AGN components, including SMBH, accretion disk, BLR, torus and EELR (or NLR). Additionally, a near-circular orbiting SMO (red) embedded in the disk is also depicted.
Panel (b): The phase of recently faded AGN, when the accretion has ceased and thus the accretion disk and BLR have disappeared while the torus and EELR at a larger scale remain there.
Panel (c): The TDE phase, depicting the disrupted star (yellow) approaching the SMBH on an inclined orbit (green line).
Panel (d): The QPE phase, in which each collision between the SMO and the disk formed by the TDE produces an X-ray eruption.
}
\label{pic}
\end{figure*}

\section{Conclusion: a unified scenario}
\label{sec:conclusion}

As shown in the previous two sections, (recently faded) AGNs, TDEs and QPEs are strongly correlated.
We conclude this work by proposing a unified scenario that naturally accounts for the correlations among the three phenomena.
We will first briefly explain the simple physical picture of each stage in this scenario, then discuss the key physical processes involved.

As depicted in Figure~\ref{pic}, the AGN phase plays a central role in this scenario. In the standard AGN phase (panel a), both the TDE rate and the EMRI formation rate are boosted. However, observationally, the identification of TDEs in AGNs is rather complicated due to the dilution of bright AGN emission and the lack of predicted hallmark features for these special TDEs. 
As the AGN phase is off, the accretion disk has vanished and an EMRI orbiting around the SMBH at radius $r=\mathcal{O}(10^2)M_\bullet$ is left (panel b). 
A TDE then occurs during the recently faded AGN phase, when the TDE rate remains elevated and there is no challenge to identify it without AGN (panel c).
After $\mathcal{O}(1)$ years, a TDE  disk forms which is generally misaligned with respect to the EMRI, as a result, QPEs are generated from quasiperiodic collisions between the EMRI and the disk (panel d).

During the AGN phase, the AGN disk boosts the wet EMRI formation rate by either capturing SMOs from the nuclear stellar cluster \citep{Pan2021prd} or lifting the star formation rate in the disk \citep{Sigl2007,Levin2007,Derdzinski2023}.  
SMOs in the disk are expected to migrate inward driven by the excited density waves until $r=\mathcal{O}(10^2) M_\bullet$ where the migration slows down and 
gradually is taken over by GW emissions \citep{Pan2021prd, Pan2022}.
Meanwhile, the TDE rate in AGNs is also predicted to be significantly elevated, as the presence of an accretion disk can effectively alter the stellar orbits
via adding a axisymmetric perturbation to the spherical gravitational potential of the SMBH or direct collisions with the disk, allowing them to approach the loss cone of the central SMBH more efficiently~\citep{Karas2007,Kennedy2016,Kaur2025}. Note, however, that the number of reported TDEs in AGNs (e.g, \citealt{Blanchard2017,Kankare2017,Liu2020,Zhang2022}) is very limited so far, mainly because the traditional search has simply neglected flares in AGNs to avoid the trouble of distinguishing between the TDE and various types of impostors~\citep{Zabludoff2021}.
Interestingly, recent simulations suggest that the TDE rate may be even enhanced in the recently faded AGN phase, when the accumulated stars, formed in the outer regions of the accretion disk, are rapidly scattered into the loss cone as a result of the vanishing inner standard thin disk. These TDEs are easily detected and may provide an ideal explanation for the overrepresentation of TDEs in PSB galaxies~\citep{Wang2024}. All of these processes may be involved in boosting the TDE rate (and the formation rate of loss-cone EMRIs in a similar way)  during the AGN phase.

In the near future, the discovery rate of TDEs is expected to increase steadily thanks to the advanced optical time-domain surveys such as the Rubin Observatory Legacy Survey of Space and Time (LSST, \citealt{LSST2019}) and the Wide Field Survey Telescope (WFST, \citealt{Wang2023WFST}). On the other hand, the QPEs can be detected during the X-ray follow-up observations of these TDEs similar to the case of AT~2019qiz, or independently identified through X-ray surveys like the Einstein Probe~\citep{Yuan2025EP}. This offers a promising prospect to verify the unified scenario based on a much larger sample of QPE-TDE associations.

\begin{acknowledgments}
We thank the referee for very positive and constructive
comments, which have improved the manuscript significantly.
This work is supported by the National SKA Program of China (2022SKA0130102), the National Natural Science Foundation of China (grants 12192221,12393814), the Strategic Priority Research Program of the Chinese Academy of Sciences (XDB0550200) and the China Manned Space Project. 
\end{acknowledgments}

%




\bibliography{ref}{}
\bibliographystyle{aasjournal}

\end{document}